\documentclass[pra,aps,twocolumn,superscriptaddress,floatfix,noshowpacs,nofootinbib,10pt]{revtex4-2}

\usepackage[utf8]{inputenc}     	
\usepackage{amsmath, amssymb}
\usepackage[colorlinks=true]{hyperref}
\usepackage{color}
\usepackage{graphicx}
\usepackage[normalem]{ulem}

\usepackage{tikz}

\usepackage{braket}

\usepackage{mathtools}
\usepackage{stmaryrd}
\usepackage{float}
\usepackage{textalpha}

\begin{document}
\title{Subdiffusion in wave packets with periodically kicked interactions}

\author{Cl\'ement Duval}
\email{clement.duval@lkb.upmc.fr}
\affiliation{Laboratoire Kastler Brossel, Sorbonne Universit\'{e}, CNRS, ENS-PSL Research University, 
Coll\`{e}ge de France; 4 Place Jussieu, 75005 Paris, France}

\author{Dominique Delande}
\email{dominique.delande@lkb.upmc.fr}
\affiliation{Laboratoire Kastler Brossel, Sorbonne Universit\'{e}, CNRS, ENS-PSL Research University, 
Coll\`{e}ge de France; 4 Place Jussieu, 75005 Paris, France}

\author{Nicolas Cherroret}
\email{nicolas.cherroret@lkb.upmc.fr}
\affiliation{Laboratoire Kastler Brossel, Sorbonne Universit\'{e}, CNRS, ENS-PSL Research University, 
Coll\`{e}ge de France; 4 Place Jussieu, 75005 Paris, France}

\begin{abstract}
We study the quantum dynamics of a peculiar driven system, a Bose gas subjected to periodically-kicked interactions. In the limit of infinitely short kicks, this system was recently shown to exhibit a fast exponential spreading of the wave function. Here we examine this problem for kicks or arbitrary duration and  show that, in this case, the spreading is not exponential but rather subdiffusive at long time. This phenomenon stems from the competition between the kinetic and interaction energies within the kicks, which is absent in the limit of delta kicks. Our analysis further shows that the breakdown of exponential spreading occurs at relatively short times even for extremely short kicks, suggesting that, in practice, subdiffusion should be more the rule than the exception in this system.
\end{abstract}
\maketitle

\section{Introduction} 


Advances in nonequilibrium quantum physics have recently revealed the richness of periodically-driven systems \cite{Eckardt2017, Moessner2017, Oka2019}, which in general ``heat'' to an infinite-temperature state due to the interplay between external forcing and interactions \cite{Reitter2017, Dalessio2014, Ponte2015}.
Among driven systems,  quantum kicked rotors have played a central role, as paradigmatic models for quantum chaos \cite{Casati1979, Grempel84} exhibiting the phenomenon of dynamical localization in momentum space \cite{Moore95, Chabe2008, Hainaut17, Hainaut18}. 
In the quantum kicked rotor, the role of interactions has also been explored on the basis of the Gross-Pitaevskii equation \cite{Shepelyansky1993, Gligoric2011, Cherroret14, Lellouch2020, Haldar2021}. It has been found, in particular, that even a weak nonlinearity may have a dramatic impact on the dynamics, by breaking the localization of wave packets and leading to a subdiffusive spreading. A similar phenomenon was also pointed out in nonlinear, spatially disordered chains \cite{Pikovsky2008, Kopidakis2008, Shepelyansky2009, Bodyfelt2011, Skokos2013, Cherroret14, Cherroret16}.

In the recent years, there has also been  a growing interest in driven quantum systems involving temporally-modulated interactions  \cite{Wang2009, Gong2009, Greschner14, Meinert2016}, used for instance to design synthetic gauge fields or to modify many-body quantum transport. In cold-atom experiments,  modulations of the interaction are typically implemented using the technique of Feshbach resonances \cite{Chin2010}. 
In the present work, we explore the quantum dynamics of a Bose-Einstein condensate subjected to periodically-kicked interactions. In the case of infinitely short (delta) kicks, it has been recently shown that this system exhibits an ultrafast, exponential spreading of the wave function in momentum space \cite{Zhao2016, Guarneri2017}. Such an exponential spreading was confirmed by methods of classical chaos based on the calculation of Lyapunov exponents \cite{Guarneri2017} and on a mapping with a generalized kicked rotor \cite{Zhao2016}. 
Here, we revisit this problem by considering interaction kicks or arbitrary duration. While we recover the exponential spreading in the limit of delta kicks, we find that, as soon as the kicks are finite, the spreading is no longer exponential but rather subdiffusive at long time. 
This phenomenon stems from the competition between the kinetic and interaction energies within the kicks, which   is absent in the limit of delta kicks. 
At the microscopic level, we interpret this subdiffusive spreading in terms of a mechanism of  incoherent coupling of the momentum sites. 
As regards the momentum distribution of the Bose gas, we find that the periodically-kicked interactions first give rise to an early-time exponential depletion of the condensate mode, quickly followed by the emergence of a ``thermal'' background of particles spreading subdiffusively.
Our analysis finally shows that the time scale where exponential spreading breaks down scales logarithmically with the kick duration. This indicates that, as soon as the kicks are finite, the subdiffusive motion tends to take over the exponential spreading at relatively short times, even if extremely short kicks are considered.

The manuscript is organized as follows. In Sec. \ref{sec_model}, we present our model of a Bose gas subjected to a sequence of interaction kicks. The time evolution of a wave packet in the limit of delta kicks is then presented in Sec. \ref{Sec:deltakicks}, and the results of previous works are recalled. In Sec. \ref{sec:subdiffusion}, we consider kicks of finite duration and show that the wave packet spreads subdiffusively in that case. A simple model for subdiffusion is introduced. In Secs. \ref{sec:condensate_fraction} and \ref{sec_momentum}, we then discuss how the subdiffusive motion shows up in the condensate fraction and the momentum distribution. We finally summarize and discuss our results in Sec. \ref{sec:summary}. Technical details are collected in four appendices.

\section{The model} 
\label{sec_model}


We study the mean-field, dynamical evolution of a one-dimensional Bose gas with a time-dependent interaction potential. This dynamics is described by the Gross-Pitaevskii equation
\begin{equation}
\label{eq_GPE}
i\hbar  \partial_\mathsf{t} \Psi(\mathsf{x},\mathsf{t}) = \frac{\hat{\mathsf{p}}^2}{2m} \Psi(\mathsf{x},\mathsf{t}) + {g}(\mathsf{t}) |\Psi(\mathsf{x},\mathsf{t})|^2\Psi(\mathsf{x},\mathsf{t}),
\end{equation}
with normalization $\int d\mathsf{x} |\Psi(\mathsf{x},\mathsf{t})|^2=1$ for the wave function $\Psi(\mathsf{x},\mathsf{t})$. The momentum operator is $\hat{\mathsf{p}}=-i\hbar\partial_\mathsf{x}$.
We consider a periodic, temporal modulation of the interaction term taking the form of a sequence of square pulses (or ``kicks'') of period $T$, width $\delta t$ and amplitude $gN$, with $g$ the interaction parameter and $N$ the total number of atoms:
\begin{equation}
\label{eq_gamma_def}
{g}(\mathsf{t}) =
\begin{cases}
0 & \text{if $\mathsf{t} \in \left[n T, (n+1)T - \delta t \right]$,}~ ~  n \in \mathbb{Z},\\
gN & \text{otherwise.}
\end{cases}  
\end{equation}
In practice, such a sequence can be realized by applying a periodic magnetic field modulation to the atomic cloud, exploiting a Feshbach resonance.
From now on, we denote by $L$ the system size and assume periodic boundary conditions, thus describing a ring geometry. This implies that the eigenstates $p$ of the momentum operator are quantized in units of $2\pi\hbar q/L$, where $q \in \mathbb{Z}$ is an integer.


To study 
the time evolution entailed by the sequence (\ref{eq_gamma_def}), it is convenient to work with a dimensionless version of Eq. (\ref{eq_GPE}). 
To this aim, we first rescale position, time and wave function according to
\begin{equation}
\label{eq_rescaling}
t= \mathsf{t}/T,\ \ {x}= 2\pi\mathsf{x}/L, \ \ |\psi|^2=|\Psi|^2L/2\pi,
\end{equation}
and introduce the effective Planck constant
\begin{equation}
\label{eq_hbareff}
\hbar_\text{eff}=\frac{\hbar T}{m}\left(\frac{2\pi}{L}\right)^2.
\end{equation}
This leads to the dimensionless Gross-Pitaevskii equation
\begin{equation}
\label{eq_GPE_dimensioneless}
i \hbar_\text{eff} \partial_t \psi(x,t) = \frac{\hbar_\text{eff}^2\hat{q}^2}{2}\psi(x,t) +  \gamma(t) |\psi(x,t)|^2\psi(x,t),
\end{equation}
where the reduced momentum operator is $\hat{q}=-i\partial_x$. The dimensionless position $x$ lies in the interval $[0,2\pi)$, and the new wave function still obeys $\int dx |\psi(x,t)|^2=1$. 
The dimensionless, self-interaction modulation is now given by
\begin{equation}
\label{eq_gamma_dimensionless}
\gamma(t) =
\begin{cases}
0 & \text{if $t \in \left[n, n+1 - \delta t/T \right]$,}~ ~  n \in \mathbb{Z},\\
\gamma& \text{otherwise,}
\end{cases}  
\end{equation}
where $\gamma=2\pi gN\hbar_\text{eff}T/L\hbar$.

In \cite{Zhao2016, Guarneri2017}, Eqs. (\ref{eq_GPE_dimensioneless}) and (\ref{eq_gamma_dimensionless}) were investigated in the limit of Dirac delta kicks, i.e., for $\delta t/T \to0$, $\gamma\to\infty$ 
with the product  $\gamma \delta t/T$ constant: $\gamma(t)= \gamma \delta t/T\sum_n\delta(t-n)$. With this model, which is known as the Gross-Pitaevskii map \cite{Guarneri2017}, the authors of \cite{Zhao2016, Guarneri2017} observed a strongly chaotic dynamics characterized by an exponential spreading of the wave function in momentum space. In \cite{Richter2018}, this model was also shown to support stroboscopic solitonic solutions.
A particular consequence of taking the limit of pure delta kicks is that the kinetic energy is irrelevant at the specific times where the kicks are nonzero. 
This is no longer the case as soon as the kick duration is finite: during the kicks, the 
kinetic energy cannot be neglected and competes with the interaction term.  This is precisely the situation we explore in the following, where we will show that this competition dramatically modifies the spreading of wave packets.

The time evolution of the state vector during one period (free evolution and kick) is governed by the evolution operator
\begin{eqnarray} \label{evol_op}
\quad\mathcal{\hat{U}}(n)&= \mathcal{T}
&\exp\left[-i\! \displaystyle\int_{n\!-\!\delta t/T}^{n}\!\!\!dt' \bigg(\frac{\hbar_\text{eff}\hat{q}^2}{2}+\frac{\gamma|\psi|^2}{\hbar_\text{eff}}\bigg)
\right]\nonumber\\
&\times& \exp\left[
-i\! \displaystyle\int_{n\!-\!1}^{n\!-\!\delta t/T}
\!\!\!dt' \frac{\hbar_\text{eff}\hat{q}^2}{2}
\right],
\end{eqnarray}
where $\mathcal{T}$ is the time-ordering operator. In this expression, the first exponential refers to the evolution during kick $n$, while the second one describes the free evolution stage before it. 
To study the system's dynamics, from now on we focus for simplicity on the limit $\hbar_\text{eff}\gg 1$, 
excluding quantum resonances where $\hbar_{\text{eff}}$ is a rational multiple of $4 \pi$ \cite{Lepers2008}. Therefore, the phase $\sim \hbar_\text{eff}$ accumulated during the free evolution stage is very large, such that it can be accurately replaced by a random variable $\phi_q$ uniformly distributed  over $\left[0, 2 \pi\right]$.  Note that we cannot apply the same random phase approximation for the kinetic phase in the first exponential of Eq. (\ref{evol_op}), which is of the order of the product $\hbar_\text{eff}\delta t /T$, not necessarily large.
To deal with the latter, it is convenient to introduce 
the change of variables $s(n)=(T/\delta t)t'+n(1-T/\delta t)$, so that 
\begin{equation} \label{evol_op2}
\quad\mathcal{\hat{U}}(n)\!=\! \mathcal{T}
\exp\!\left[-i\! \displaystyle\int_{n\!-\!1}^{n}\!\!\!ds\bigg(\frac{\hat{q}^2}{2 f^2}+\gamma^*|\psi|^2\bigg)
\right]
\!\exp(
-i\phi_q),
\end{equation}
where 
\begin{equation}
\gamma^*=\frac{\gamma\delta t}{T\hbar_\text{eff}}=\frac{2\pi g N\delta t}{L\hbar}
\end{equation}
is the effective interaction strength, and
\begin{equation}
f=\sqrt{\frac{T}{\delta t \hbar_\text{eff}}}=\frac{L}{2 \pi}\sqrt{\frac{m}{\delta t \hbar}}
\end{equation}
controls the amplitude of the kinetic energy during the kicks. Information about the finite duration of the kicks is entirely contained in this parameter. In particular, when $f=\infty$ the kinetic term in Eq. (\ref{evol_op2}) vanishes and one effectively recovers the delta-kick limit of \cite{Zhao2016, Guarneri2017}. 

The evolution operator  (\ref{evol_op2}) can be readily implemented numerically to describe the dynamics entailed by Eq. (\ref{eq_GPE_dimensioneless}) for arbitrary kick durations. 
To this aim, we 
introduce the Fourier transform
\begin{equation}
\psi_q(t) = \frac{1}{\sqrt{2\pi }} \int_0^{2\pi}\! \! dx \, e^{-i q x} \psi(x,t).
\end{equation}
Recalling that  permissible reduced momenta $q \in \mathbb{Z}$, this relation can be inverted as
\begin{equation}
\psi(x, t) = \frac{1}{\sqrt{2\pi }} \sum_q\! \! \, e^{i q x} \psi_q(t),
\end{equation}
with the normalization $\sum_q|\psi_q(t)|^2=1$. In the following, we take as an initial state the wave function
\begin{equation}\label{ini_state}
\psi_q(t=0) \propto \exp(-\lambda^2 q^2),
\end{equation}
of momentum width $\lambda^{-1}$ typically smaller than 1. In practice, this state is a good model for the narrow momentum distribution of a Bose-Einstein condensate. Note that the corresponding spatial distribution is broad, nearly uniform at the scale of the system size, and it remains uniform on average during the time evolution.
In contrast, we will see in the next section that its momentum distribution exhibits a non-trivial behavior as a result of the combined effect of the nonlinear and kinetic terms in Eq. (\ref{evol_op2}). 

In order to numerically describe the evolution of the wave packet $\psi_q(t)$, we successively apply the evolution operator (\ref{evol_op2}) to the initial state (\ref{ini_state}), using 
 a second-order split-step method to evaluate the wave function at each time step $\Delta s$ \cite{Weideman1986}. The latter is always chosen much smaller than the unit time scale, typically $\Delta s=1/500$. 
In the simulations we discretize the interval $[0,2\pi]$ into $N_s$ spatial steps, where $N_s\gg1$. 
All our results, finally, are averaged over typically $N_r \sim 10^4$ realizations of the random phase $\phi_q$. 
Some observables of interest, like ${|\psi_0|^2}$, are however very sensitive to the numerical instability inherent to the non-linear Schr\"odinger equation \cite{Lakoba2012, Semenova2020}. These instabilities are discussed  in more detail in Appendix \ref{numerics_detail}. In order to circumvent them, we have worked with a high-precision arithmetic whenever exponential sensitivity to initial conditions was the limiting factor. Typically our algorithm ensured $N_d = 100$ significant decimal digits. We have always checked that increasing $1/\Delta s$, $N_s$ or $N_d$ does not alter our numerical calculations.
\section{Wave-packet spreading for delta-kicks}
\label{Sec:deltakicks}

\begin{figure}
\includegraphics[scale=0.7]{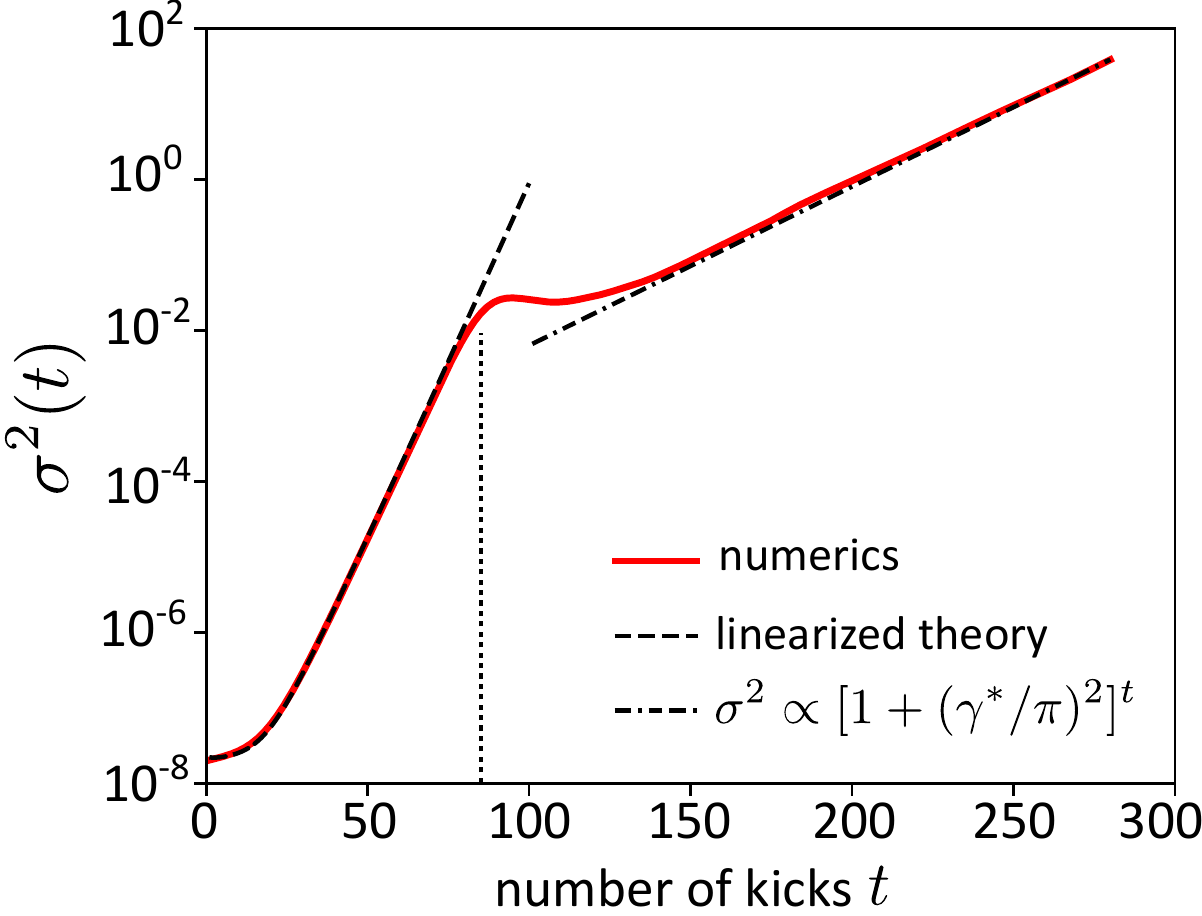}
\caption{Mean-square width of the wave packet as a function of the number of kicks $t$, for $f=\infty$ (delta kicks). Here $\gamma^*=0.7$, $\lambda=3.03$. The dashed curve is the theoretical prediction for short times, Eq. (\ref{eq_psi1_t}), and the dashed-dotted curve is the prediction of \cite{Zhao2016, Guarneri2017} for long times, Eq. (\ref{Guarneri_result}). The vertical dotted line indicates the position of the characteristic time $t_E$ separating the two regimes of exponential growth, given by Eq. (\ref{eq_t1}).
\label{fig_sigma2}}
\end{figure}

To characterize  temporal spreading of the wave packet (\ref{ini_state}) subjected to the interaction kicks, we examine the temporal evolution of its mean-square width in momentum space,
\begin{equation}
\label{Ec_expr}
\sigma^2(t)=  \sum_q q^2 \, \overline{|\psi_q(t)|^2},
\end{equation}
where the overbar refers to averaging over the random phase $\phi_q$ accumulated between the kicks. 
In this section, we focus on the limit $f=\infty$ of delta kicks.
The corresponding time evolution of $\sigma^2(t)$ is shown in Fig. \ref{fig_sigma2}, and emphasizes two
distinct dynamical regimes.
 $\sigma^2(t)$ first grows exponentially up to  a certain characteristic time $t_E$ ($t_E\simeq 80$ in Fig. \ref{fig_sigma2}). Then, for $t>t_E$, the growth slows down albeit it remains exponential.

To clarify the origin of this result, we first discuss the short times  $t<t_E$. 
The growth of $\sigma^2(t)$ in this regime corresponds to a fast initial depletion of the condensate from $q=0$ to the neighboring momentum sites $q\ne0$. To describe it quantitatively, we start from the evolution equation of the Fourier modes during a given kick $n$:
\begin{equation}
i\partial_s \psi_q =\
\frac{\gamma^*}{2\pi}\sum_{q_1,q_2}\psi_{q_1}^*\psi_{q_2}\psi_{q+q_1-q_2}.
\label{eq_GPE_q}
\end{equation}
This equation corresponds to the first exponential term in the evolution operator $\mathcal{\hat{U}}(n)$. Recall that $s$ is the variable introduced in Eq. (\ref{evol_op2}), labeling continuous times during a non-linear pulse over the range $\left]n-1, n  \right[$.
 At short time, mostly modes $q=-1,0,1$ are populated. Neglecting the other modes and assuming $|\psi_j|^2\ll|\psi_0|^2$ ($j=\pm1$), we can linearize Eq. (\ref{eq_GPE_q}), which leads to $\psi_0(s)\simeq \psi_0(n-1) \exp[-i \gamma^* (s-n+1)/2\pi]$ and
\begin{equation}
i\partial_s \psi_j\simeq 
\frac{\gamma^*}{\pi}
\psi_j+\frac{\gamma^*}{2\pi} \psi_0^2\psi_{-j}^*.
\label{eq_GPE_1}
\end{equation}
These equations still contain nonlinear factors that are conveniently removed with the gauge transformation $\tilde\psi_{j}=\psi_{j} / \psi_0 $.  Then we introduce
the circular state vector for the first Fourier mode after the kick $n$, $\Gamma(n)=(\Re\, \tilde\psi_1(n),\Im\, \tilde\psi_1(n))^\intercal$, and assume for simplicity  $\tilde\psi_1=\tilde\psi_{-1}$ \cite{footnote}. The propagation of this state vector over one period obeys  $\Gamma(n)=U\ \Gamma(n-1)$, 
where the transfer matrix $U$ is given by
\begin{equation}
\label{eq_transfermatrix}
U\simeq\begin{pmatrix}
1 & 0\\
-\gamma^*/\pi & 1
\end{pmatrix}
\begin{pmatrix}
\cos\phi_1 & -\sin\phi_1\\
\sin\phi_1 & \cos\phi_1
\end{pmatrix}.
\end{equation}
The second matrix in the right-hand side describes the free-space propagation between two interacting pulses, which involves  the uniformly distributed kinetic phase $\phi_1$, see Eq. (\ref{evol_op2}). The  first matrix, on the other hand, describes the propagation during the kick $n$, and is inferred from Eq. (\ref{eq_GPE_1}) and its complex-conjugated version. The population of the first Fourier mode after $n$ kicks, finally, follows from:
\begin{equation}
\label{eq_psi1square}
\overline{|\psi_1(t=n)|^2}=\frac{1}{2\pi}\int_0^{2\pi}d \phi_1\, \lVert U^n\Gamma(0) \rVert^2.
\end{equation}
The calculation of the integral over $\phi_1$ is detailed in Appendix \ref{label_appendix1}.  At weak interaction strength $\gamma^*/2\pi\ll1$, we find
\begin{equation}
\label{eq_psi1_t}
\overline{|\psi_1(t)|^2}\simeq e^{-2\lambda^2}
\left[1
+\frac{1}{2\pi}\sqrt{\frac{\gamma^*}{2t}}e^{\gamma^*t/\pi}
\right].
\end{equation}
This  indeed describes an exponential growth of the first Fourier mode, at the rate $ \gamma^*/\pi$. 
In the short-time regime where Eq. (\ref{eq_psi1_t}) is valid, the mean-square width (\ref{Ec_expr}) is dominated by the contribution of the first mode, i.e., $\sigma^2(t)\simeq 2\overline{|\psi_1(t)|^2}$. This result is shown in Fig. \ref{fig_sigma2} (dashed curve), and matches very well the numerical simulations.

The approach above can also be used to access $t_E$, by noting that at $t\sim t_E$ the linearization breaks down. Precisely, we find $t_E$ from the condition $\text{max}_{\phi_1}\lVert U^{t_E}\Gamma(0) \rVert^2 =  1/2$, i.e. from the specific configuration
where $|\psi_1|^2 $ is maximal and becomes of the order of $|\psi_0|^2$ (recall that, as long as only two modes are considered, $|\psi_0|^2+2|\psi_1|^2$=1  due to the normalization). 
This criterion leads to (see Appendix \ref{label_appendix1})
\begin{equation}
\label{eq_t1}
t_E\simeq \frac{2\pi\lambda^2}{\gamma^*}.
\end{equation}
$t_E$ is the typical time needed for the wave function to spread over the portion of phase space spanned by the $q=-1,0,1$ states, the so-called Ehrenfest time~\cite{Haake}. We will use this terminology from now on.
Notice that, the smaller $\lambda$ -- that it when the initial state is broader -- the shorter the Ehrenfest time.
Equation (\ref{eq_t1}) is reported in Fig. \ref{fig_sigma2}, and it correctly describes the crossover to the long-time expansion regime. 

At $t\sim t_E$, the modes $|q|>1$ turn out to significantly impact the dynamics, so that the effectively occupied phase space grows, and a different dynamics is observed. Study of the regime $t> t_E$ was the main goal of the previous works \cite{Zhao2016, Guarneri2017}, where it was found that
\begin{equation}
\sigma^2(t)\sim\exp[t\ln(1+(\gamma^*/\pi)^2)].
\label{Guarneri_result}
\end{equation}
In \cite{Guarneri2017}, in particular, the authors derived
this exponential growth by rewriting Eq. (\ref{eq_GPE_dimensioneless}) in the form of a generalized kicked-rotor model and by studying the evolution of $\sigma^2(t)$ in the corresponding classical map. A similar exponential growth was also found in \cite{Mieck2005}, in a slightly different model involving a linear kicking potential on top of the nonlinear sequence of kicks. The exponential law (\ref{Guarneri_result}) is shown in Fig. \ref{fig_sigma2}, and well captures the numerical results at long time.

\section{Finite kicks: long-time  subdiffusion}
\label{sec:subdiffusion}
\begin{figure}
\includegraphics[scale=0.57]{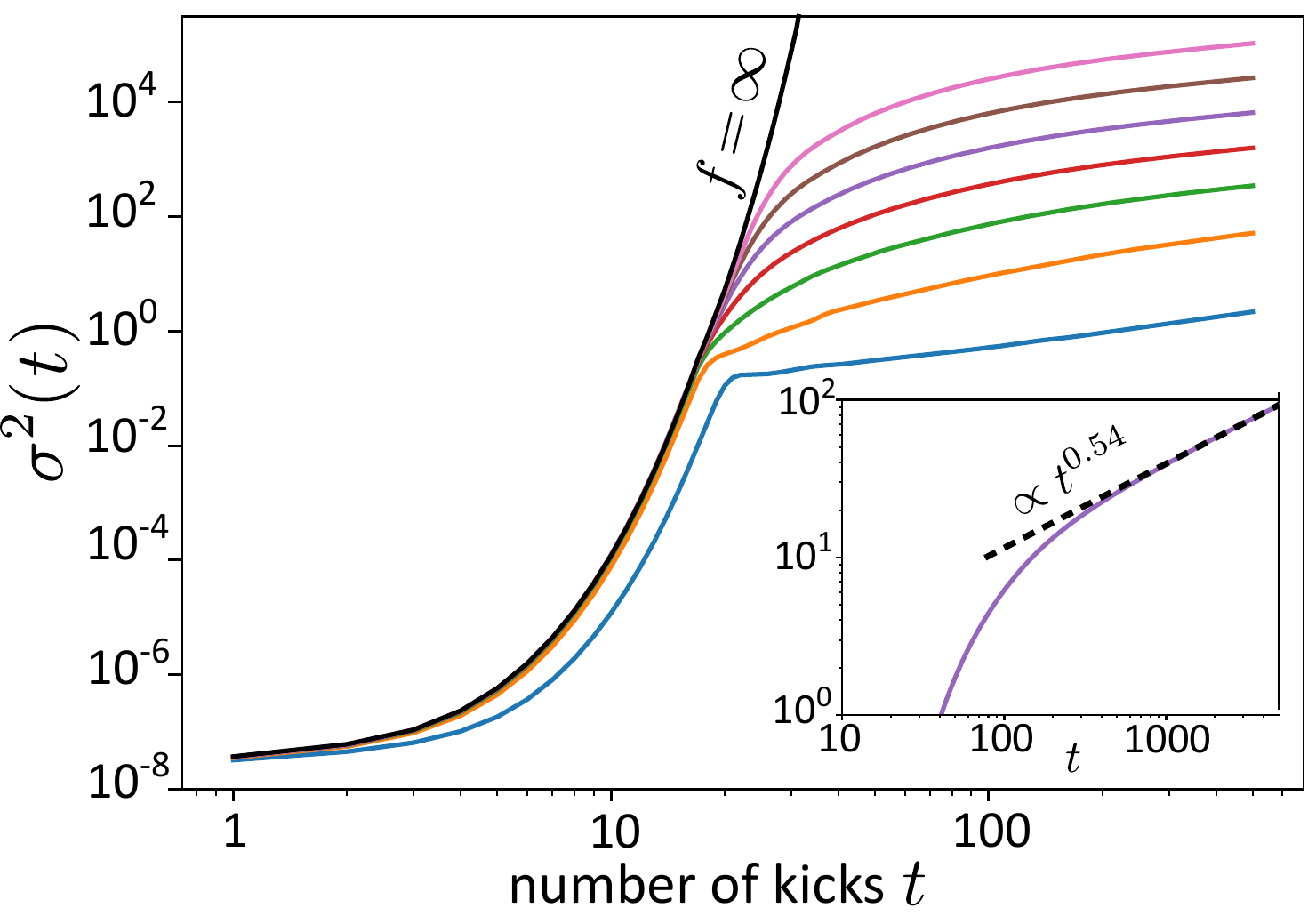}
\caption{Main panel: mean-square width of the wave packet as a function of the number of kicks $t$ up to $t=600$, for finite, increasing values of $f$. Solid curves from bottom to top correspond to $f=1,2,4,8,16,32,64$. 
Here $\gamma^*=4$, $\lambda=3.03$. The black curve corresponds to numerical results in the limit $f=\infty$ (delta kicks). 
Inset: mean-square width up to $t=5000$ at $f=16$, emphasizing the subdiffusive behavior. 
\label{Fig2_Sigma_vs_time_long}}
\end{figure}

Let us now consider the case of finite kicks, which is the main object of the paper.
The evolution of the wave-packet mean-square width for finite values of $f$ is displayed in Fig. \ref{Fig2_Sigma_vs_time_long}. 
The limit $f=\infty$ is also shown for comparison (black curve). 
The figure shows that when $f$ is finite, the behavior of $\sigma^2(t)$ at times $t<t_E$ remains well captured by Eq. (\ref{eq_psi1_t}), except, perhaps, for small values of $f$. On the other hand, a dramatically different evolution emerges beyond $t_E$: the growth of $\sigma^2(t)$ is no longer exponential but rather algebraic, with a prefactor  increasing with $f$. 
An analysis up to $t\simeq 5\times 10^3$ is shown in the inset, and suggests the following algebraic scaling at long time:
\begin{equation}
\label{eq_subdiffusion}
\sigma^2(t)\propto t^{1/2}.
\end{equation} 


Such a subdiffusive behavior can be understood in terms of a mechanism of ``heating'' where the spreading wave packet is incoherently coupled to its neighboring sites via the nonlinearity. A similar mechanism was shown to also take place in the context of the nonlinear Schr\"odinger equation in the presence of disorder \cite{Shepelyansky1993, Gligoric2011}.
To clarify this idea, we start from the evolution equation during a kick at finite $f$:
\begin{equation}
i\partial_s \psi_q=\frac{q^2}{2 f^2}\psi_q+
\frac{\gamma^*}{2\pi}\sum_{q_1,q_2}\psi_{q_1}^*\psi_{q_2}\psi_{q+q_1-q_2}.
\label{eq_GPE_q_f}
\end{equation}
Let us now consider a certain momentum-site $q$ located at the border of the spreading wave packet. 
At the contact with the wave packet, the amplitude of this site evolves according to Eq. (\ref{eq_GPE_q_f}). 
\begin{figure}
\includegraphics[scale=0.65]{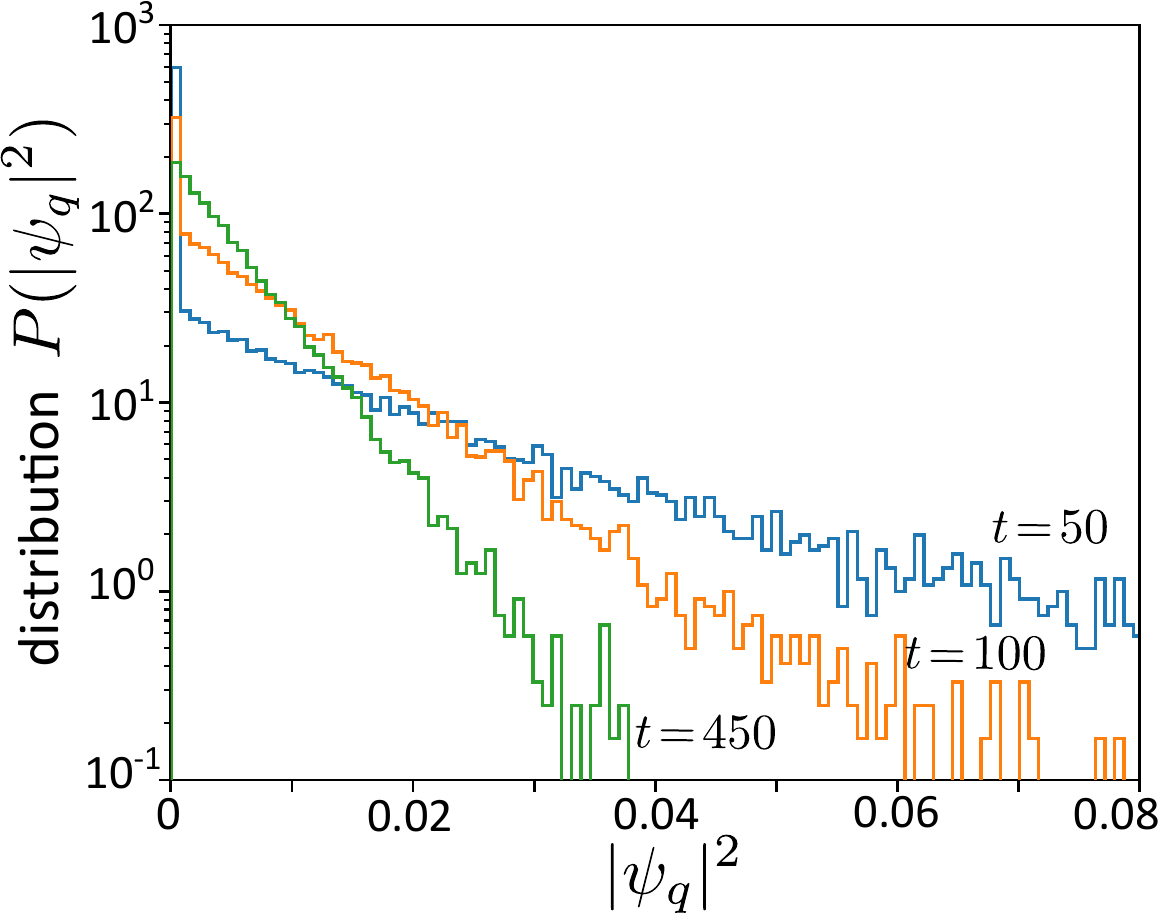}
\caption{Probability distribution of the momentum density,  $P(|\psi_q(t)|^2)$, at different times and fixed $f=16$ and $\gamma^*=4$. Here we choose $q=2$ as an example, the distributions at other $q$-values behaving similarly. At long time, the distribution becomes exponential, indicating that the $\psi_q$ are purely complex random Gaussian variables.
\label{Fig_distributions}}
\end{figure}
We then make the hypothesis that the coupling between this site and the spreading wave packet consists in an incoherent heating mechanism where the second term in the right-hand side of Eq. (\ref{eq_GPE_q_f}) is replaced by a random noise. 
This assumption is motivated by the fact that the complex amplitudes $\psi_q$ become Gaussian random variables at long enough time, as is confirmed by the numerical simulations in Fig. \ref{Fig_distributions}.
We thus replace Eq. (\ref{eq_GPE_q_f}) by the Langevin equation:
\begin{equation}
i\partial_s \psi_q\simeq\frac{q^2}{2 f^2}\psi_q+\frac{\gamma^*}{2\pi}f^{2}\rho(s)^{3/2} \eta(s),
\label{eq_Langevin}
\end{equation}
where $\rho(s)$ is the momentum density of the spreading wave packet, which for the simplicity of the argument we here take uniform, and $\eta(s)$ is an uncorrelated random noise, satisfying $\overline{\eta(s)\eta(s')}=\delta(s-s')$.
The prefactor $f^2=\mathcal{N}^2$ stems from the number $\mathcal{N}$ of terms effectively involved in each sum in the right-hand side of Eq. (\ref{eq_GPE_q_f}).  
At long time, this number 
must be related to the strength of the kinetic term in Eq. (\ref{eq_GPE_q_f}) which is precisely responsible for the existence of subdiffusion. As a rough estimate, we identify $\mathcal{N}$ with the typical value of $q$ for which the kinetic term in Eq. (\ref{eq_GPE_q_f}) is of the order of $1$, i.e., $q\sim \mathcal{N}\sim f$. This scaling is also in good agreement with a more quantitative analysis of the inverse participation ratio discussed in Appendix \ref{appendix_fluctuations}.

Solving the Langevin equation  provides the average squared amplitude at the site $q$ at late time $t \gg t_E$, $\overline{|\psi_q(t)|^2}$. The result is  $\overline{|\psi_q(t)|^2}\sim f^4\gamma^{*2}\rho^3 t$ \cite{Pomeau2017}. 
From this, we infer that the typical time $t_s$ it takes for the wave packet to ``heat'' the site $q$ is such that $\rho\sim f^4\gamma^{*2}\rho^3 t_s$, giving $t_s^{-1}\sim f^4\gamma^{*2}\rho^2$. Finally, we assume that the wave-packet spreading can be described by a nonlinear diffusion equation of the type $d\sigma^2(t)/dt= D(t)$, with a diffusion coefficient $D(t)$ proportional to the heating rate $t_s^{-1}$. This leads to
\begin{equation}
\label{eq_sigma2_diffusion}
\sigma^2(t)\sim f^4 \gamma^{*2}\rho^2 t.
\end{equation}
For a uniform wave packet, $\rho(t)=1/\sigma(t)$ due to norm conservation. The solution of Eq. (\ref{eq_sigma2_diffusion}) for $\sigma^2(t)$ then immediately yields
\begin{equation}
\sigma^2(t)\sim f^2\gamma^* t^{1/2},
\end{equation}
which reproduces the time evolution in Fig. \ref{Fig2_Sigma_vs_time_long}. We have also verified that the scaling of the prefactor in $f^2\gamma^*$ is well reproduced by the numerical simulations at long time.

\section{Condensate fraction and crossover to the delta-kick limit}
\label{sec:condensate_fraction}

\begin{figure}
\includegraphics[scale=0.61]{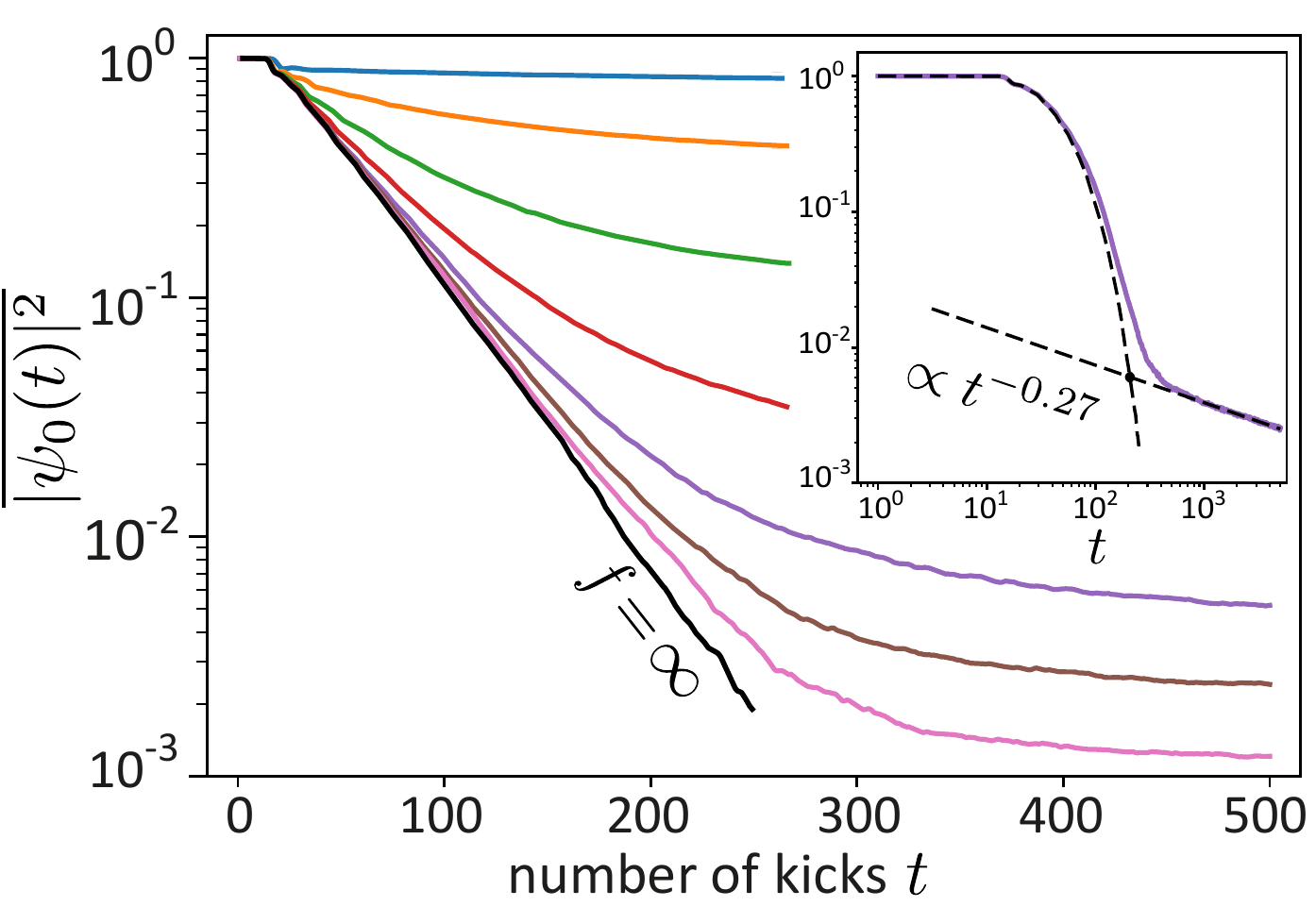}
\caption{Average condensate fraction $\overline{|\psi_0(t)|^2}$ versus time for increasing values of $f$. 
Solid curves from top to bottom correspond to $f=1,2,4,8,16,32,64$, and the black line is the $f=\infty$ limit.
Here $\gamma^*=4$ and $\lambda=3.03$. The inset shows $\overline{|\psi_0(t)|^2}$ in double logarithmic scale for $f=16$. Interpolating between the short-time (exponential) and  long-time (algebraic) decays provides an estimate of the time scale $t_f$ beyond which the description in terms of delta kicks becomes incorrect.
\label{Fig_DecayBEC}}
\end{figure}

\begin{figure}
\includegraphics[scale=0.65]{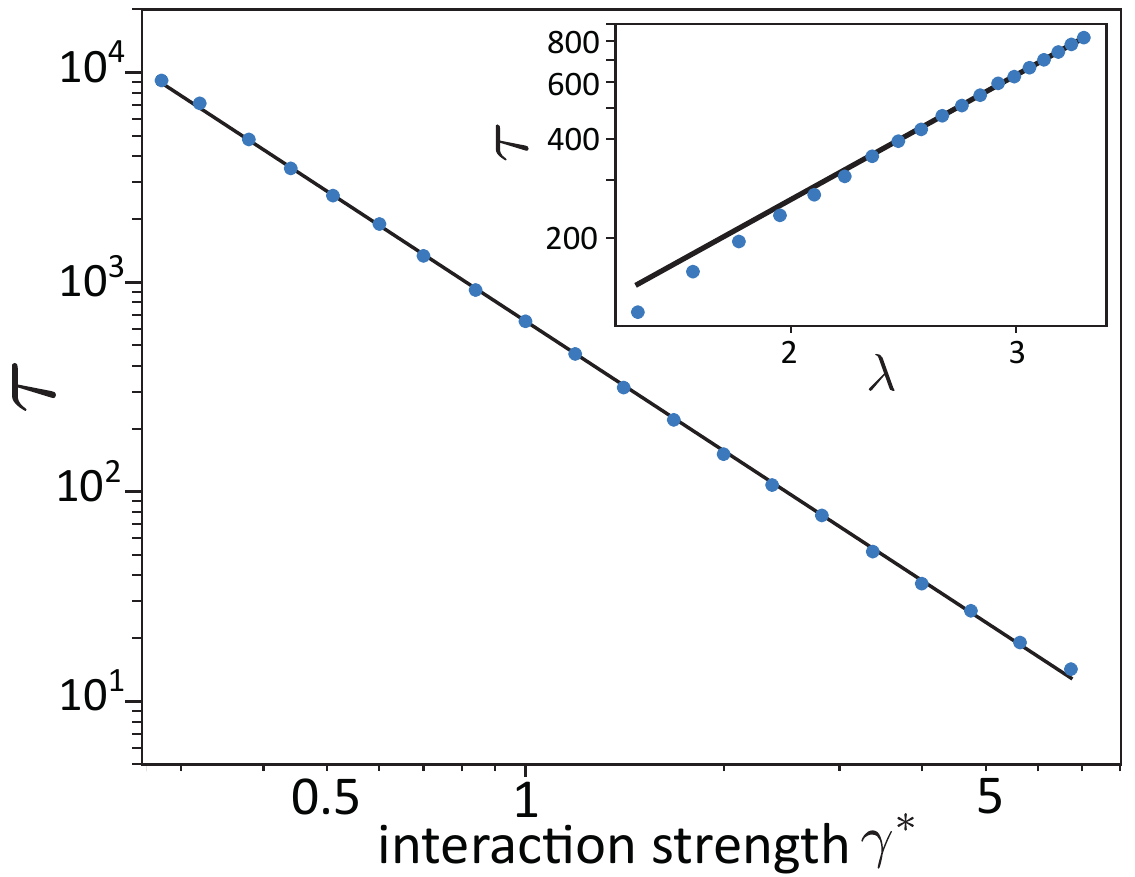}
\caption{Characteristic time governing the exponential decay $\overline{|\psi_0(t)|^2}\sim \exp[-(t-t_E)/\tau]$ of the average condensate fraction in the limit $f= \infty$, extracted from numerical simulations. The main panel shows $\tau $ as a function of the interaction strength $\gamma^*$, at fixed $\lambda=3.03$. Dots are numerical data and the solid curve is a linear fit providing $\tau\propto1/\gamma^{*2.05}$. The inset shows $\tau $ as a function of $\lambda$, at fixed $\gamma^*=1$. The solid curve is a linear fit giving $\tau\propto\lambda^{2.16}$. Together, the plots provide $\tau \simeq 70.8 \lambda^{2.16} / {\gamma^*}^{2.05}$, in very good agreement with the theoretical prediction (\ref{eqtau}).
\label{Fig_tau}}
\end{figure}

Another relevant quantity for probing the difference between finite and delta kicks is the average Bose condensate fraction, defined as $\smash{\overline{|\psi_0(t)|^2}}$. This quantity is shown in the main panel of Fig. \ref{Fig_DecayBEC}.
Like for $\sigma^2(t)$, we first discuss the case of delta kicks, $f=\infty$, which is represented by the black curve. Numerically, we find that in this limit the condensate fraction decays exponentially from $t_E$: $\smash{\overline{|\psi_0(t)|^2}}\simeq \exp[-(t-t_E)/\tau]$. 
To understand this behavior and access the time scale $\tau$, we note that $\smash{\overline{|\psi_0(t)|^2}}$ is essentially the probability for the condensate mode to remain populated at time $t$. 
This probability is governed by specific realizations of the random phases $(\phi_1, \phi_2, \dots)$ for which the small-$q$ modes have not grown exponentially at a time $t$ larger than $t_E$. 
Precisely, in Appendix \ref{appendix_gen_tE} we show that 
when $\psi_q$ grows exponentially while the populations of the first $q-1$ modes do not, the condensate gets depleted at the typical time $t_q = q t_E$; The probability of such an event is $(1 - \mathcal{P})^{q-1}$,
where $\mathcal{P}= \gamma^*/\pi^2\ll 1$. In other words,
\begin{eqnarray}\label{exp_law_cond}
\overline{|\psi_0(t_q)|^2}&\simeq&(1-\mathcal{P})^{q-1}\simeq \exp[-(q-1)\mathcal{P}]\nonumber\\
&=&\exp[-\gamma^*(t_q-t_E)/(\pi^2 t_E)].
\end{eqnarray}
This result confirms the exponential decay observed numerically, and identifies the characteristic time $\tau$ to deplete the condensate as
\begin{equation}
\label{eqtau}
\tau\simeq\frac{\pi^2 t_E}{\gamma^*}=\frac{2\pi^3\lambda^2}{\gamma^{*2}}.
\end{equation}
We have also studied this time scale numerically as a function of $\gamma^*$  and $\lambda$, as shown by the two plots in Fig. \ref{Fig_tau}, and the results agree with the prediction (\ref{eqtau}).

Let us now consider kicks of finite duration. The time evolution of the condensate fraction in that case is illustrated by the colored curves in Fig. \ref{Fig_DecayBEC}.
As for $\sigma^2$, as soon as $f$ is finite we observe a clear deviation from the exponential scaling, $\overline{|\psi_0(t)|^2}$ decreasing much more slowly. 
An analysis of the condensate fraction over longer times, shown in the inset of Fig. \ref{Fig_DecayBEC}, again points toward a subdiffusive behavior at finite $f$. We find $\overline{|\psi_0(t)|^2}\sim 1/t^{1/4}$, which is fully consistent with the subdiffusive law (\ref{eq_subdiffusion}) for the mean-square width (see also Sec. \ref{sec_momentum}).

The time evolutions of the condensate fraction at finite and infinite $f$  discussed above can be used to estimate the characteristic time $t_f$ beyond which the model $f=\infty$ of delta kicks can no longer be reliably utilized to describe the dynamics. This question is crucial from a practical point of view, since in a real experiment the duration of the  kicks cannot be made arbitrarily small, especially if the bosonic interactions are modulated using Feshbach resonances. To find $t_f$, we interpolate the temporal scalings of the condensate fraction at finite and infinite $f$, where $\smash{\overline{|\psi_0(t)|^2}\sim 1/(f t^{1/4})}$ and $\smash{\overline{|\psi_0(t)|^2}\sim\exp[-(t-t_E)/\tau]}$, respectively. This method, illustrated in the inset of Fig. \ref{Fig_DecayBEC}, yields
\begin{equation} \label{tf_expr}
t_f\sim \tau\ln f\sim \frac{2\pi^3\lambda^2}{\gamma^{*2}}\ln f.
\end{equation}
The logarithmic   dependence of $t_f$ on $f$ has a remarkable consequence. Even for extremely large values of $f$, i.e., for extremely short kick durations, the breakdown of the exponential decay of the condensate fraction [or the exponential growth of $\sigma^2(t)$] occurs at relatively short times (this phenomenon is, in fact, visible by eye in Fig. \ref{Fig2_Sigma_vs_time_long}). This implies that, in a real experiment that unavoidably involves finite kicks, the subdiffusive behavior described in the present work should be more the rule than the exception.

\section{Momentum distribution}
\label{sec_momentum}

\begin{figure}
\includegraphics[scale=0.45]{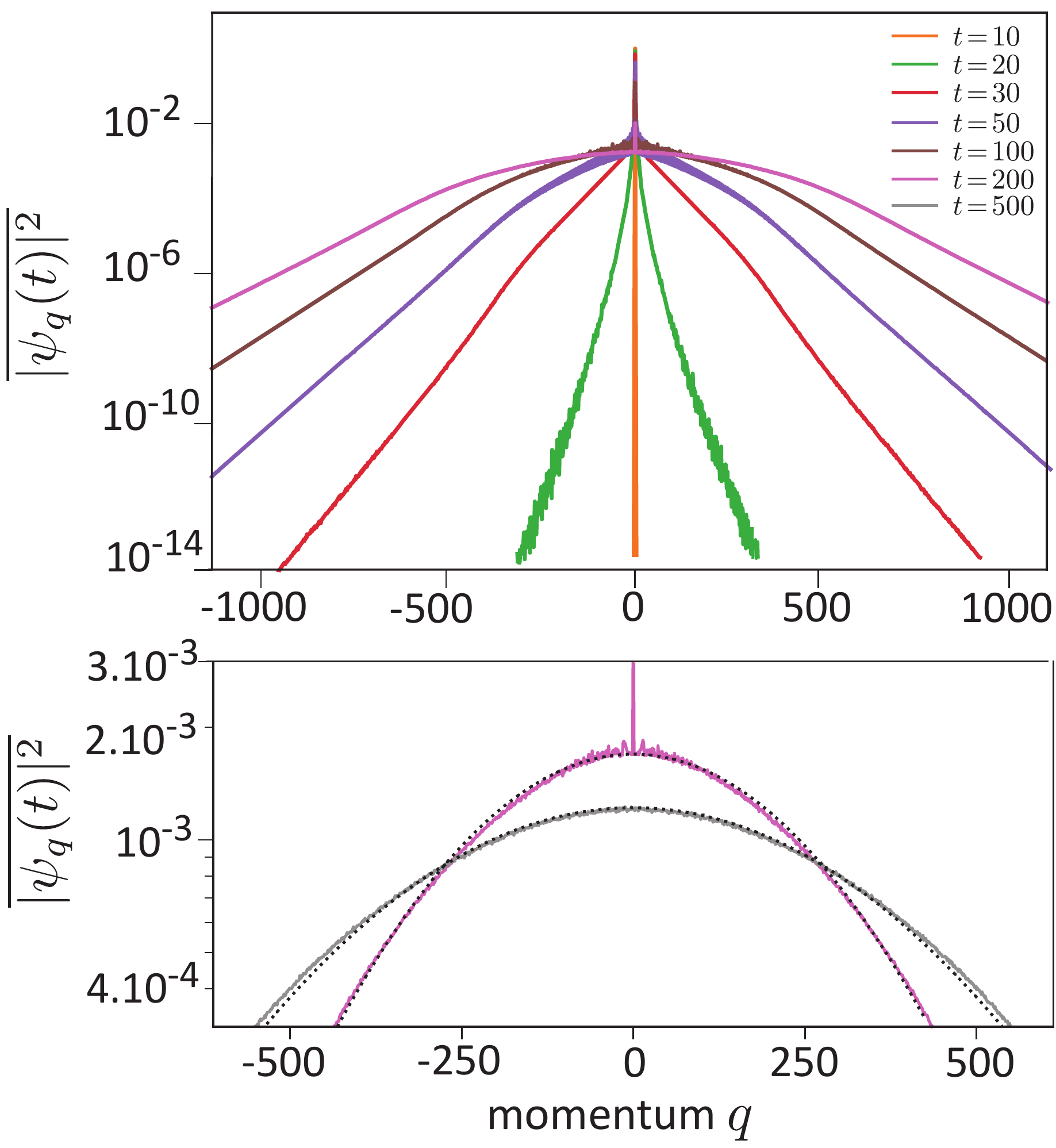}
\caption{
Upper panel: average momentum distribution of the wave packet at increasing times, at fixed $f=64$, $\gamma^*=4$ and $\lambda=3.03$. At short time, the condensate mode $\overline{|\psi_0(t)|^2}$ decays exponentially at the scale of $\tau$. This decay is accompanied by a slow growth of the thermal modes $q\ne 0$. The latter control the subdiffusive evolution of the wave-packet mean-square width according to Eq. (\ref{eq_subdiffusion}).
The lower panel is a zoom on the central part of the distribution at $t=200$ and $t=500$. These profiles are very well described by the Gaussian distribution (\ref{eq:gaussian_profile}), without any fit parameter.
\label{Fig_momdensity}}
\end{figure}

All the above findings can be summarized by looking at the average momentum distribution of the Bose gas at different times. Such distributions are displayed in the upper panel of Fig. \ref{Fig_momdensity}.
The distributions first exhibit an exponential decay of the condensate mode, $\overline{|\psi_0(t)|^2}$, at the scale of $\tau$, quickly accompanied by a slow growth of the ``thermal'' modes $q\ne 0$. The latter control the subdiffusive evolution of the wave-packet mean-square width according to Eq. (\ref{eq_subdiffusion}).
The lower panel of the figure is a zoom on the central part of the distribution at $t=200$ and $t=500$. As soon as the condensate fraction is negligible, i.e., at times of the order of a few $\tau$, we numerically find that this central part is always very well approximated by the (normalized) Gaussian profile 
\begin{equation}
\label{eq:gaussian_profile}
\overline{|\psi_q(t)|^2}\simeq\frac{1}{\sqrt{2\pi\sigma^2(t)}} \exp\left[  -  \frac{q^2}{2\sigma^2(t)}   \right],
\end{equation}
where $\sigma^2(t)$ is the mean-square width of the \textit{whole} distribution [satisfying Eq. (\ref{eq_subdiffusion})].
Note that this law in particular implies $\smash{\overline{|\psi_0(t)|^2}\sim 1/t^{1/4}}$, in accordance with the results of the previous section. 
This Gaussian shape is another marked difference with the behavior observed in the $f=\infty$ limit, for which the profile is known to be exponential at all momenta \cite{Mieck2005, Guarneri2017, Zhao2016}. 

\begin{figure}
\includegraphics[scale=0.65]{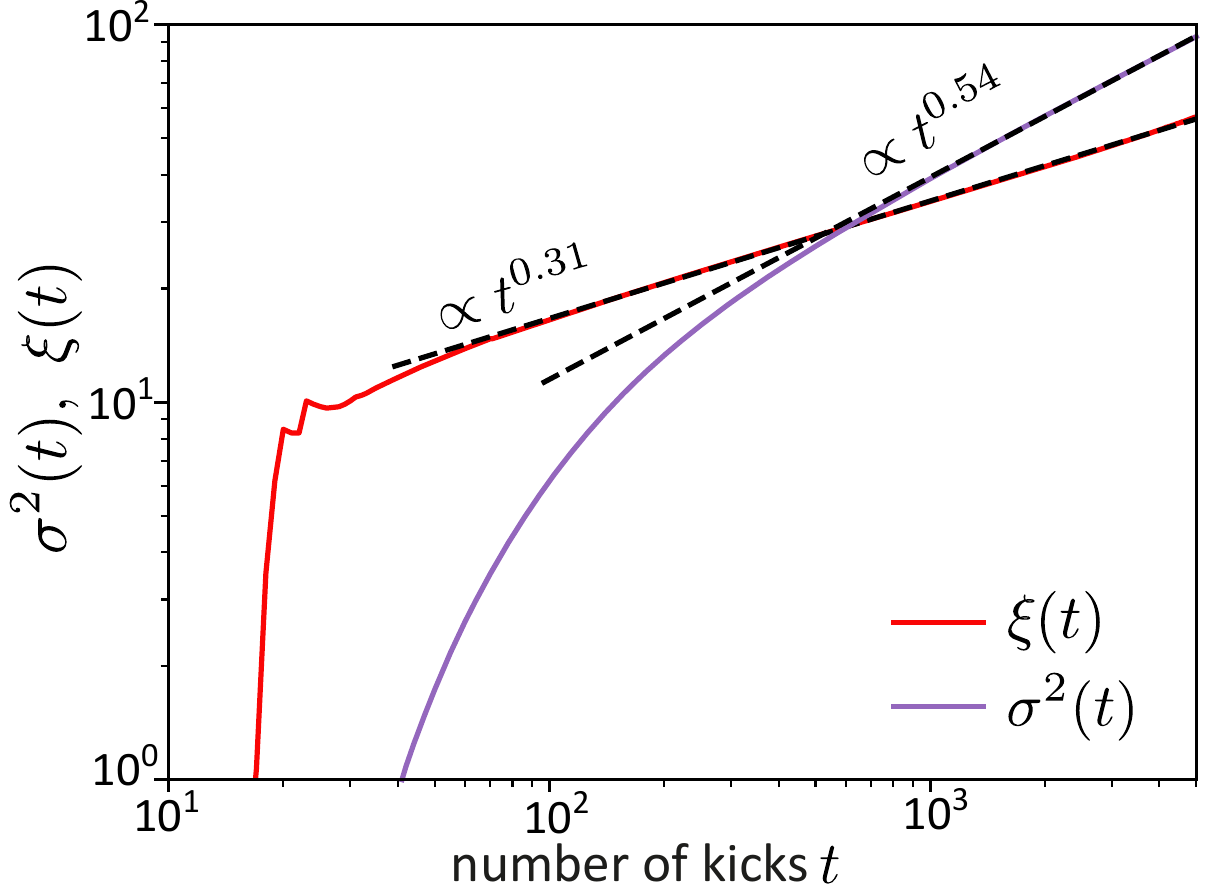}
\caption{
Time evolution of the exponential decay length $\xi(t)$ governing the far wings of the momentum distribution. 
A linear fit (dashed line) suggests an algebraic scaling close to $\xi(t)\sim t^{1/3}$. The time evolution of the variance $\sigma^2(t)$ is shown for comparison. Here $f=16$, $\gamma^*=4$ and $\lambda=3.03$.
\label{Fig_xisub}}
\end{figure}

When $f$ is finite, nevertheless, our numerics suggests that the far $q$-wings of the momentum distribution also decay exponentially, see Fig. \ref{Fig_momdensity}. 
At variance with the $f=\infty$ limit, however, here the exponential decay length $\xi(t)$ does not grow exponentially in time, but rather subdiffusively. The numerical results of Fig. \ref{Fig_xisub}, indeed, suggest $\smash{\overline{|\psi_q(t)|^2}\sim \exp[-|q|/\xi(t)]}$ with $\xi(t)\sim t^\alpha$ and $\alpha$ close to ${1/3}$. 
Although the degree of universality of this value is not yet clear to us, 
this subdiffusive law is apparently different from the one governing the variance $\sigma^2(t)$, see Fig. \ref{Fig_xisub}. There is, of course, no contradiction at this stage, since the exponential wings  provide a negligible contribution to the variance of the whole distribution. While we have not been able to find an analytical basis for the scaling of the far wings, it could stem from a mechanism different from the incoherent heating  discussed in Sec. \ref{sec:subdiffusion}, involving, for example, resonant coherent coupling between the spreading wave packet. Clarifying this question would constitute an interesting challenge for future work.

\section{Discussion and summary}
\label{sec:summary}

By considering a Bose gas subjected to a periodic modulation of the interactions taking the form of finite kicks, we have found evidence for the emergence of a mechanism of subdiffusive spreading of the wave function beyond a characteristic Ehrenfest time. This result has to be contrasted with the case of delta kicks, where the spreading is always exponential. 
We have interpreted the subdiffusive motion in terms of an incoherent heating process for the nonlinear coupling of momentum sites. 
Beyond this analysis, however, one may ask what fundamental mechanism could explain the different dynamics observed for finite and delta kicks. A possible explanation could be the different nature of the  quantities conserved within a given kick in the two scenarios. 
Indeed, when $f=\infty$, the evolution equation during one kick can be immediately integrated to yield
\begin{equation} \label{sol_GPmap}
	\psi(x,n+1) = e^{-i \gamma^*|\psi(x,n)|^2} \psi(x,n).
\end{equation}
The norm of the wave function is thus conserved for all point $x$ in that case. This local constraint in position space suggests that, conversely, the coupling between modes is  weakly constrained in momentum space, resulting in a very fast  spreading of the wave packet. 
In contrast, when $f$ is finite, such a local solution no longer exists and, instead, the nonlinear Sch\"odinger equation involves only \emph{global} integrals of motions of the form $\int dx \, \psi^*(x,t) Q_j(x,t)$ with the $Q_j(x,t)$ defined via recursion relations \cite{Magri1978}. We expect this global character to translate into a much weaker coupling between the modes in reciprocal space.

Our analysis has also revealed that, for finite kicks, the subdiffusive motion takes over the exponential spreading at a characteristic time that scales logarithmically with the kick duration. This characteristic time is thus always relatively short, even in the limit of the extremely short kicks. This suggests that, in practice, subdiffusion rather than exponential spreading of wave packets should be more naturally observed in this system.

\appendix

\section{Exponential growth of the first Fourier mode}
\label{label_appendix1}

In this appendix, we calculate the population of the first Fourier mode at short time,
\begin{equation}
\label{eq_psi1square2}
\overline{|\psi_1(t=n)|^2}=\frac{1}{2\pi}\int_0^{2\pi}d \phi_1\, \lVert U^n\Gamma(0) \rVert^2,
\end{equation}
where the transfer matrix $U$ is given by Eq. (\ref{eq_transfermatrix}), and the initial state vector is $\smash{\Gamma(0)=(e^{-\lambda^2},0)^\intercal}$. By symmetry, the contribution of negative $\phi_1$ equals the one of positive $\phi_1$, so that the integral average in Eq. (\ref{eq_psi1square2}) can be replaced by $1/\pi\int_0^\pi d\phi_1$. Explicitly, the matrix $U$ reads
\begin{equation}
\label{transfert}
U  = \begin{pmatrix}
\cos\phi_1 & -\sin\phi_1 \\
-\frac{\gamma^*}{\pi}\cos\phi_1\!+\!\sin\phi_1  & \cos\phi_1\!+\!\frac{\gamma^*}{\pi} \sin\phi_1
\end{pmatrix},
\end{equation}
whose eigenvalues are of the form $\mu\pm\sqrt{\mu^2-1}$, with $\mu=\cos\phi_1+(\gamma^*/2\pi)\sin\phi_1$. 
For values of $\phi_1$ such that $\mu^2 - 1 < 0$, the spectrum of $U$ is unimodular.
The exponential growth of the first Fourier mode observed in the numerical simulations, on the other hand, stems from the contributions of $\phi_1$ such that $\mu^2-1>0$.
Indeed, in this case one of the two (distinct) eigenvalues is of modulus strictly larger than one, eventually leading to an exponential growth of $\overline{|\psi_1(t)|^2}$. 
For weak interactions $\gamma^*/2\pi\ll1$, $\mu^2-1\simeq \phi_1(\gamma^*/\pi-\phi_1)$ such that the values of $\phi_1$ leading to 
$\mu^2-1>0$ lie in the interval $[0,\gamma^*/\pi]$. Note that this upper bound also defines the probability $\mathcal{P}=\gamma^*/\pi^2$ that the first excitation grows exponentially when one ``draws'' a value of $\phi_1$.

The diagonalization of $U$ provides
\begin{equation}
\label{transfert_2}
U^n\Gamma(0)  = -\exp(-\lambda^2)\begin{pmatrix}
\frac{\eta}{x}\sinh x n -\cosh x n  \\
\frac{\nu}{x}\sinh x n
\end{pmatrix},
\end{equation}
for $\gamma^*\ll1$, where $\eta= \phi_1 \gamma^*/2\pi$, $x=\sqrt{\phi_1(\gamma^*/\pi-\phi_1)}$ and $\nu=\gamma^*/\pi-\phi_1$. At leading order in $\gamma^*$, this leads to
\begin{eqnarray}
\label{eq_psi1square_bis}
&&\overline{|\psi_1(t=n)|^2}
\simeq e^{-2\lambda^2}+\\
&&\ \ \ \ \frac{\gamma^* e^{-2\lambda^2}}{\pi^2}\int_0^{\gamma^*/\pi}
\frac{d \phi_1}{\phi_1} \sinh^2[t\sqrt{\phi_1(\gamma^*/\pi-\phi_1)}]\nonumber.
\end{eqnarray}
The integral over $\phi_1$ can be calculated by a saddle point approximation, the saddle point being $\phi_1=\gamma^*/2\pi$. This gives
\begin{equation}
\overline{|\psi_1(t)|^2}\simeq e^{-2\lambda^2}
\left[1
+\frac{1}{2\pi}\sqrt{\frac{\gamma^*}{2t}}e^{\gamma^*t/\pi}
\right],
\label{eq_psi1_exp}
\end{equation}
which is Eq. (\ref{eq_psi1_t}) of the main text. 
To find the Ehrenfest time (\ref{eq_t1}), finally, we simply apply the criterion given in the main text, $\text{max}_{\phi_1} \lVert U^{t_E}\Gamma(0) \rVert^2 =  1/2$, together with Eq. (\ref{eq_psi1square_bis}), noting that the maximum of the integrand is attained at the saddle point $\phi_1=\gamma^*/2\pi$.

\section{Exponential depletion of the condensate}
\label{appendix_gen_tE}

\begin{figure}[H]\centering
	\begin{tikzpicture}
		\node (blabla) at (0,0){
			\includegraphics[scale=0.6]{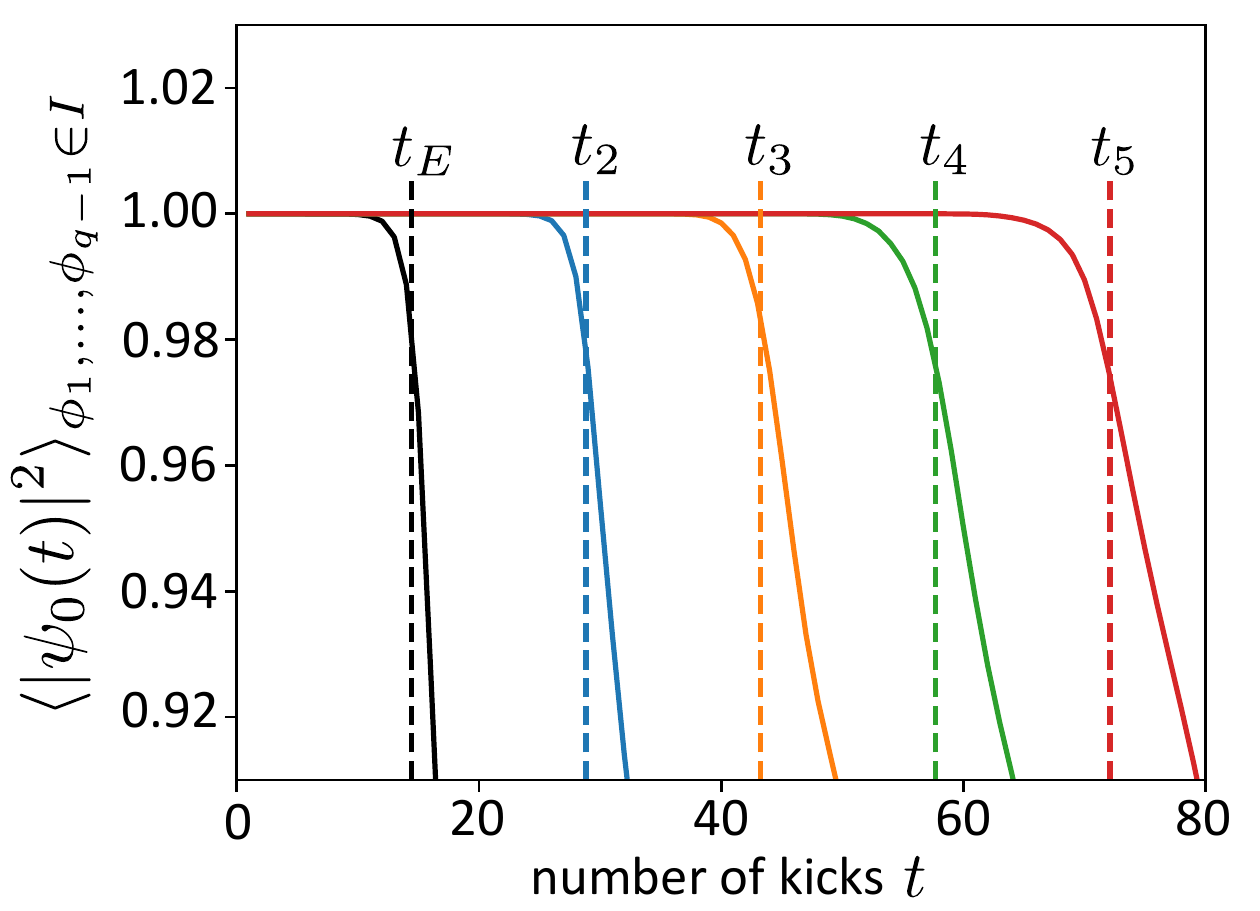}
		};
	\end{tikzpicture}
	\caption{
Condensed fraction $\langle | \psi_0 |^2  \rangle_{\phi_1, \dots, \phi_{q-1} \in I}$ for $q=1, \dots, 5$ from left to right. These curves are numerically obtained by averaging the population $|\psi_0|^2$ over the random vector $(\phi_1, \phi_2 \dots)$ under the constraint $\phi_1, \dots, \phi_{q-1} \in I$. The case $q=1$ (black curve) coincides with the unrestricted average $\overline{|\psi_0|^2}$.  Here $f = 64$, $\lambda = 3.03$ and $\gamma^*=4$. Vertical dashed lines indicate the positions of the theoretical predictions (\ref{gen_tE}) for $t_q$.
}\label{Pap_graphe24}
\end{figure}

In this appendix, we examine realizations of the random phases $(\phi_1, \phi_2, \dots)$ for which some small-$q$ modes grow exponentially and reach a significant population at a time $t$ larger than $t_E$. As mentioned in the main text, these configurations govern the exponential decay  of the condensate fraction.
Below we analyze such realizations and show that 
when $\psi_q$ grows exponentially while the populations of the first $q-1$ modes do not, the condensate gets depleted around the time $t_q = q t_E$, the probability of such an event being $(1 - \mathcal{P})^{q-1}$ with $\mathcal{P}= \gamma^*/\pi^2\ll 1$. 
To this aim, we start by considering realizations of  $\phi_1$ and $\phi_2$  for which the population $|\psi_2|^2$  grows exponentially but the population $|\psi_1|^2$ does not. According to the analysis of Appendix \ref{label_appendix1}, $|\psi_1|^2$ does not grow exponentially when $\phi_1\in I = \, [ \gamma^*/\pi, \pi ]$ and, for weak interactions, this event occurs with the probability $1-\mathcal{P} \simeq 1$.
In that case, the spectrum of the transfer matrix $U$ is unimodular, so that $\psi_1$ rotates in time in the complex plane. The depletion of $|\psi_0|^2$ is then mainly controlled by the behavior of the second excitation. Our aim is then to find the new time scale $t_2$ at which the linearization breaks down due to the exponential growth of $|\psi_2|^2$. 
Similarly to Appendix \ref{label_appendix1}, this can be achieved by linearizing the Gross-Pitaevskii equation (\ref{eq_GPE_q_f}) keeping the modes $q=0,1,2$ only, and identifying $t_2$ as the time where the linearization procedure breaks down.
Let us call $\Gamma_2$ the state vector of the second mode, $\Gamma_2=(\Re\, \tilde\psi_2,\Im\, \tilde\psi_2)^\intercal$, where $\tilde\psi_2 = \psi_2 / \psi_0 $. Linearization of the equation of motion (\ref{eq_GPE_q_f}) during a pulse gives
\begin{equation}\label{eq_gamma2}
 	\partial_s \Gamma_2 =  \begin{pmatrix}
		0& 0 \\
		- \gamma^*/\pi & 0
	\end{pmatrix} \Gamma_2 + \frac{\gamma^*}{2 \pi}\begin{pmatrix}
		\Im \, \tilde{\psi}_1^2 \\
		- 2 |\tilde{\psi}_1|^2 - \Re \, \tilde{\psi}_1^2
	\end{pmatrix}. 
\end{equation}
Over one period, this can be written as
\begin{equation}
	\Gamma_2(n) = U \Gamma_2(n-1) + \Sigma(n),
\end{equation}
where $\Sigma(n)$ refers to the rightmost term of Eq. (\ref{eq_gamma2}), which implicitly depends on $\phi_1$. $U$ now denotes the transfer matrix (\ref{eq_transfermatrix}) where $\phi_1$ has been replaced
by $\phi_2$.
Using that $\smash{ \lVert U \Gamma_2(0)\rVert\sim e^{-4\lambda^2} \ll \lVert\Sigma_1\rVert \sim e^{-2\lambda^2}}$, we obtain
\begin{equation} \label{eq_gamma2_sum}
	\Gamma_2(n) = \sum_{m=0}^{n-1} U^m \Sigma(n-m),
\end{equation}
for times $n \geq 1$. Since $\tilde{\psi}_1$ rotates in the complex plane, we also infer $ \lVert\Sigma(m) \rVert \sim e^{-2\lambda^2}$ whatever $\phi_1,m$. Therefore, the same argument as in Appendix \ref{label_appendix1} shows that the exponential grows of $\langle|\psi_2(t)|^2 \rangle_{\phi_1 \in I}$ arises from realizations $\phi_2\in \bar{I}=[ 0 ,  \gamma^*/\pi ]$, occuring with probability $\mathcal{P}= \gamma^*/\pi^2$. The average population of the second mode is then mainly governed by the saddle  point of the term $U^{n-1}\Sigma(1)$ in Eq. (\ref{eq_gamma2_sum}):
\begin{equation}\label{ansatz_aq}
\langle|\psi_2(t>1)|^2\rangle_{\phi_1 \in I}\propto \exp(-4\lambda^2) \exp \left[\gamma^* (t-1) / \pi \right].
\end{equation}
Up to logarithmic corrections, the only difference with the second term in the r.h.s. of Eq. (\ref{eq_psi1_exp}) stems from the different prefactor $e^{-4\lambda^2}$ instead of $e^{-2\lambda^2}$. Accordingly, the time scale $t_2$ at which the linearization hypothesis breaks down is given by
\begin{equation}
t_2 \simeq \frac{4 \pi \lambda^2}{\gamma^*}=2t_E.
\end{equation}
This analysis can be extended to higher $q>2$: we consider events of probability $(1-\mathcal{P})^{q-1}$ for which $\phi_1, \dots, \phi_{q-1} \in I$. Then, the depletion of the condensate is mostly controlled by the $q$-th mode, and the typical time scale at which linearization breaks down is 
\begin{equation}\label{gen_tE}
	t_q \simeq q t_E.
\end{equation}
We have corroborated this prediction numerically by computing the average population $\langle|\psi_0|^2\rangle$ under the constraint $\phi_1, \dots, \phi_{q-1} \in I$, see Fig.~(\ref{Pap_graphe24}).

\section{Statistics of fluctuations}
\label{appendix_fluctuations}
\begin{figure}[H]\centering
	\begin{tikzpicture}
		\node (blabla) at (0,0){
			\includegraphics[scale=0.6, trim=0cm 0 0 0]{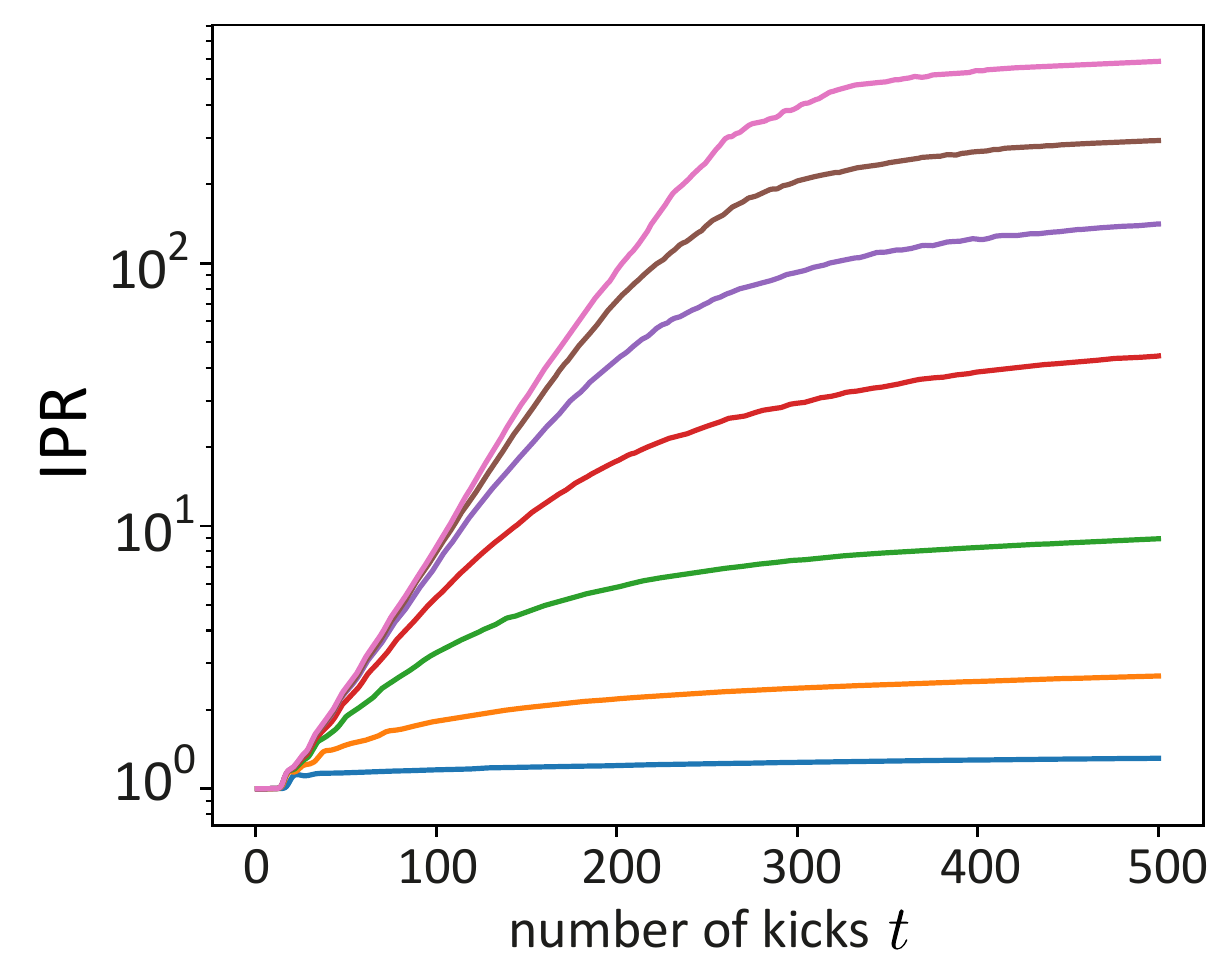}
		};
	\end{tikzpicture}
	\caption{Inverse participation ratio versus time. Solid curves from bottom to top correspond to $f=1,2,4,8,16,32,64$. Here $\gamma^*=4$ and $\lambda = 3.03$.
}\label{Fig_IPR}
\end{figure}

In addition to the dispersion $\sigma^2(t)=\sum_q q^2 \, \overline{|\psi_q(t)|^2}$, the spreading of the wave packet can be characterized by the inverse participation ratio (IPR) $1/\sum_q \overline{|\psi_q(t)|^4}$, which measures the average number of excited modes at a given time $t$.  
The IPR is shown in Fig. (\ref{Fig_IPR}) as a function of time. 
At very short times, $1/\sum_q \overline{|\psi_q|^4} \simeq 1$ since only one mode is appreciably populated. This is the regime discussed in Sec. \ref{Sec:deltakicks}. When $t \geq t_E$, many $q$ modes start to be populated and the IPR rapidly increases. At late times, finally, the increase is slowed down as the system enters the subdiffusive regime described in Sec. \ref{sec:subdiffusion}. We also note that for sufficiently large $f$, IPR $\propto f$ at a given time. This result validates the estimation of Sec. \ref{sec:subdiffusion} for the number of modes effectively participating in the second term in the right-hand side of Eq. (\ref{eq_GPE_q_f}) in the subdiffusive regime.

\begin{figure}[H]\centering
	\begin{tikzpicture}
		\node (blabla) at (0,0){
			\includegraphics[scale=0.7, trim=0cm 0 0 0]{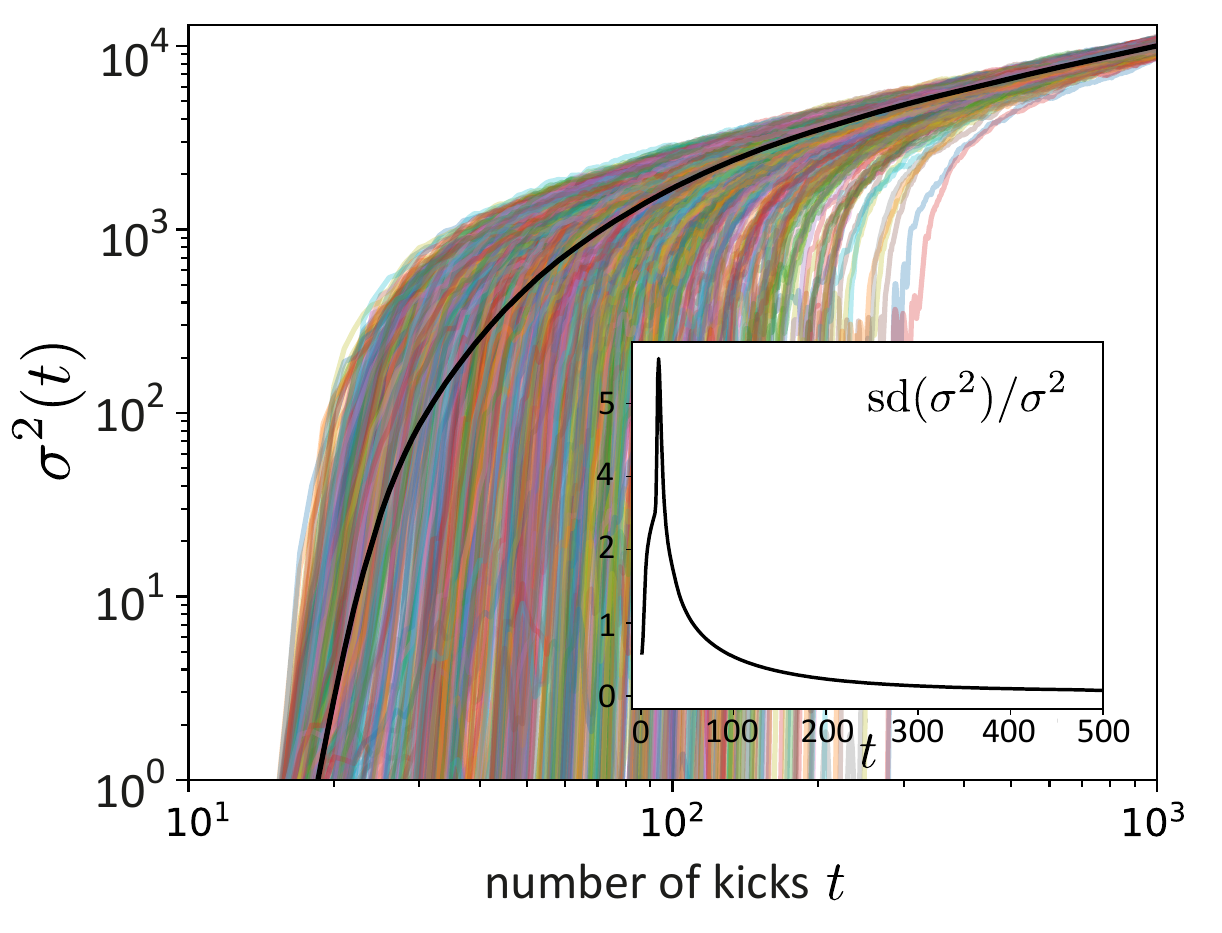}
		};
	\end{tikzpicture}
	\caption{
	Wave-packet dispersion $\sigma^2$ for $\sim 10^3$ realizations of the random phases (colored curves), and its average (thick black curve). Here $f=16$, $\lambda = 3.03$ and $\gamma^*=4$. The inset shows the standard deviation of the dispersion relative to its average.
}\label{Fig_sigmafluctuations}
\end{figure}
We finally comment on the fluctuations of the dispersion $\sigma^2=\sum_q q^2 \, |\psi_q(t)|^2$ from one realization of the random phases to the other. These fluctuations are illustrated in Fig. \ref{Fig_sigmafluctuations}, which shows the behavior of the dispersion for many realizations. The fluctuations are typically large in the vicinity of the Ehrenfest time while the dispersion becomes self-averaging in the long-time, subdiffusive regime.
The standard deviation of the dispersion, defined as
\begin{equation}
	\text{sd}(\sigma)\! =\! \Big[\overline{\Big(\sum_q q^2 \, |\psi_q(t)|^2\Big)^2}\!-\! \Big(\sum_q q^2 \, \overline{|\psi_q(t)|^2}\Big)^2\Big]^{1/2}\!\!,
\end{equation}
is shown in the inset of Fig. \ref{Fig_sigmafluctuations} and confirms this behavior.

\section{Numerical instabilities}
\label{numerics_detail}

\begin{figure}[H]\centering
	\begin{tikzpicture}
		\node (blabla) at (0,0){
			\includegraphics[scale=0.6, trim=0cm 0 0 0]{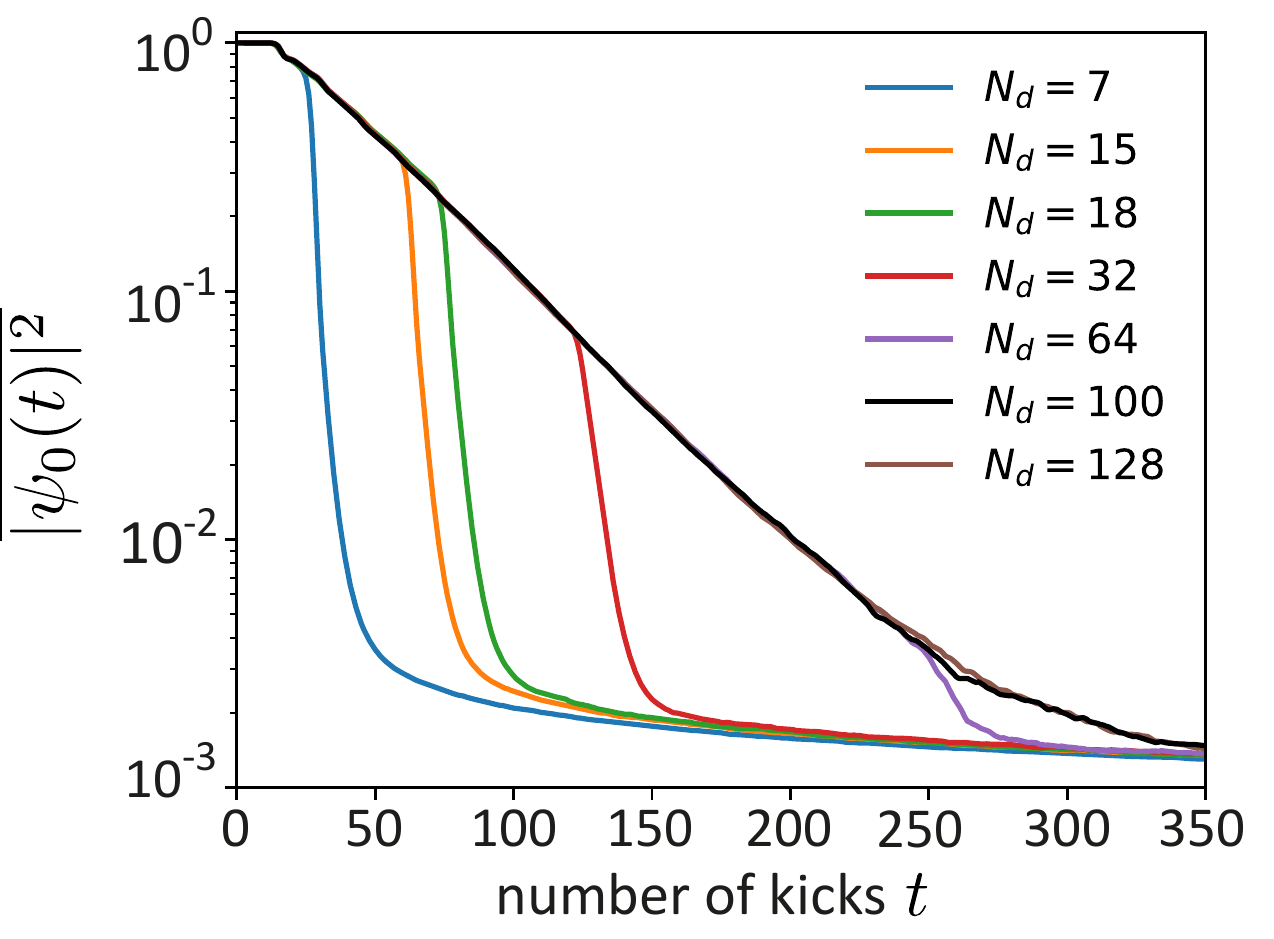}
		};
	\end{tikzpicture}
	\caption{Benchmarking of high-precision calculations using $N_d = 32, 64, 100, 128$ against results of standard fixed-precision formats (i.e. ``single'', ``double'' and ``extended double'' precisions, respectively $7,15$ and $18$ significant decimal digits). Here $f = 64$, $\lambda = 3.03$ and $\gamma^*=4$, and numerical parameters are $1/\Delta s = 500$, $N_r \sim  10^4$, $N_s = 2^{12}$.}\label{Pap_graphe7}
\end{figure}

Our numerical calculations are based on a second-order split-step method \cite{Weideman1986}. In practice, however, the specificities of our system make certain observables 
very sensitive to instabilities of this numerical scheme. 
A typical example is provided by the average condensate fraction, $\overline{|\psi_0(t)|^2}$, which decays exponentially in time. Indeed, as explained in Sec. \ref{sec:condensate_fraction} and in Appendix \ref{appendix_gen_tE}, at long time this quantity is governed by realizations of the random phases for which the populations $|\psi_{q}(t)|^2$ of other modes remain exponentially small up to large $q$ values. An accurate estimation of $\overline{|\psi_0(t)|^2}$ thus requires an accurate computation of many exponentially small $|\psi_{q}(t)|^2$, typically limited by the round-off error threshold that depends on the finite number of significant decimal digits $N_d$ representing floating-point data. 
This issue is illustrated in Fig.~\ref{Pap_graphe7}, which shows the average condensate fraction computed for increasing values of $N_d$. At too low $N_d$, the results exhibit a non-physical temporal collapse. 
In practice, for our typical choices of $ \left\{f,  \lambda, \gamma^*  \right\}$, we have found that $N_d = 100$ allows to faithfully estimate the condensate fraction up to a several hundreds kicks at all time scales.


\begin{thebibliography}{99}
\bibitem{footnote}
This assumption amounts to impose $\phi_1=\phi_{-1}$. Although this condition is not strictly verified for a given realization of the random phase, we have verified that the final result is unchanged when one considers $\phi_1\ne \phi_{-1}$. Indeed, in this case the eigenvalues of $U$, which is now a $4\times4$ matrix, are given by $e^{-i(\phi_1-\phi_{-1})/2}(\mu\pm\sqrt{\mu^2-1)}$. This is the same form as in the case $\phi_1=\phi_{-1}$, except for the additional phase factor $e^{-i(\phi_1-\phi_{-1})/2}$, which nevertheless has no impact on the final result, Eq. (\ref{eq_psi1_t}).
\end{thebibliography}

\end{document}